\pdfoutput=1
\documentclass[aps,prl,reprint,groupedaddress,floatfix]{revtex4-1}
\usepackage{blkarray}
\usepackage{amssymb}
\usepackage{amsmath}
\usepackage{epsfig} 
\usepackage{epic,eepic}

\usepackage{hyperref}
\usepackage{lipsum}
\usepackage{fancyhdr}
\usepackage{physics}
\usepackage[utf8]{inputenc}
\usepackage[vietnamese, english]{babel}
\usepackage{bbold}
\usepackage{color}
\usepackage{subfigure}

\begin{document}


\title{Effective Interactions in a Graphene Layer Induced by the Proximity to a Ferromagnet}


\author{\foreignlanguage{vietnamese}{Võ Tiến Phong}$^1$}
\email{Phong.Vo@postgrad.manchester.ac.uk} 
\author{Niels R. Walet$^1$}
\email{Niels.Walet@manchester.ac.uk}
\author{Francisco Guinea$^{1,2,3}$}
\email{Francisco.Guinea@manchester.ac.uk}
\affiliation{$^1$School of Physics and Astronomy, University of Manchester, Manchester, M13 9PY, UK}
\affiliation{$^2$Imdea Nanoscience, Faraday 9, 28015 Madrid, Spain}
\affiliation{$^3$Donostia International Physics Center, Paseo Manuel de Lardizabal 4, 20018 San Sebastian, Spain}

\date{\today}

\begin{abstract}
The proximity-induced couplings in graphene due to the vicinity of a ferromagnetic insulator are analyzed. We combine general symmetry principles and simple tight-binding descriptions to consider different orientations of the magnetization.
 We find that, in addition to a simple exchange field, a number of other terms arise. Some of these terms act as magnetic orbital couplings, and others are proximity-induced spin-orbit interactions. The couplings are of similar order of magnitude, and depend on the orientation of the magnetization. A variety of  phases, and anomalous Hall effect regimes, are possible.
\end{abstract}


\maketitle


{\it Introduction.}
Electrons in graphene placed in proximity to a ferromagnet experience an induced exchange field that modifies spin-transport properties. The exchange field results from virtual electrons hopping between the graphene layer and the ferromagnet. For instance, on an yttrium-iron garnet (YIG) substrate, this leads to magnetoresistance, spin-current-to-charge-current conversion \cite{M15}, and spin precession from induced ferromagnetism \cite{M15,LKWW17,Setal17}. Furthermore, in addition to induced ferromagnetism, we can also expect induced spin-orbit coupling (SOC) in graphene because atoms in the YIG substrate have non-negligible SOC \cite{M15}. The YIG ferromagnetism breaks time-reversal symmetry and thus gives rise to orbital couplings in the graphene layer which are not invariant under time inversion. We can expect similar effects on graphene in proximity to an EuO ferromagnet \cite{SOHRK12, WTSBS15, WLLPCCKZHH2016, S17}. Heavy atoms near a graphene layer can, in general, induce non-trivial spin-orbit couplings \cite{WHFW11}, and these effects have been experimentally confirmed \cite{COGBNB15,Wetal15,Wetal16, KOVSP17}. With such experimental platforms now available, we can attempt to engineer proximity-induced interactions in graphene for spintronics applications \cite{ZFS04, HKGF14, LSKZ14}. 

In the present work, we analyze the general interactions induced in a graphene layer by proximity to a ferromagnetic insulator with a significant SOC. We give a classification of the possible terms allowed by lattice symmetries, and present simple models which allow us to determine the relative strengths of the various terms. The resulting electronic structure of graphene depends sensitively on the balance between these couplings, and can exhibit a number of interesting topological features. We discuss the main features of the different possible phases.

\begin{figure}
\includegraphics[scale=0.16]{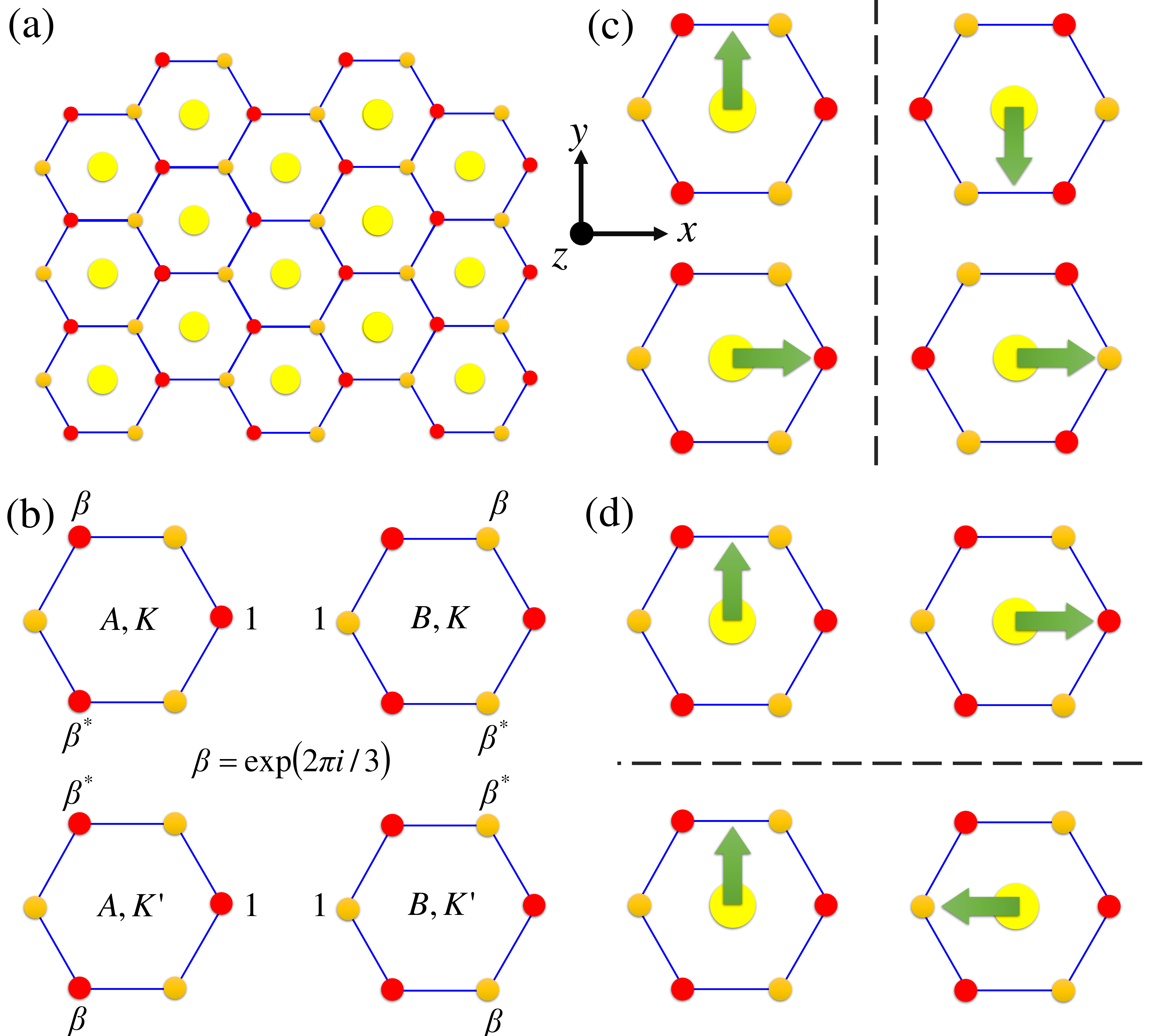}
\caption{(a) Lattice of graphene-ferromagnet bilayer. The A sublattice of graphene is shown in red, while the B sublattice is shown in orange. The yellow dots represent the ferromagnetic atoms placed on top of graphene at the center of each hexagon. (b) The phases of the Bloch wavefunction of graphene for each sublattice at the $K$ and $K'$ points. (c)-(d) Mirror reflection symmetry of in-plane magnetization. The green arrows indicate the projection direction of angular momentum, which behave as pseudovectors under reflection.}
\label{fig: lattice}
\end{figure}

{\it The model.}
For simplicity, we consider the interaction between a graphene sheet and the top layer of a magnetic substrate. While this approach does not allow us to make quantitative numerical predictions for the strength of the induced couplings in graphene, this is sufficient for a general analysis of the type of perturbations and their relative strengths. It is worth noting that the couplings between graphene and a substrate depend exponentially on the distance between the two systems, so that a quantitative theoretical study is, in any case, extremely challenging. We further assume that the magnetic atoms form a lattice commensurate with the graphene layer, and neglect the effects of disorder. This approximation implies that the allowed couplings satisfy translational symmetry. As an average translational symmetry over long scales can be defined for any substrate orientation, this approximation will capture the leading couplings. In the absence of the symmetries considered here, other terms are possible, although it can be expected that they will be of smaller magnitudes. Finally, we neglect the coupling between magnetic atoms. Such couplings will lead to dispersive bands in the graphene layer. Our goal is to define effective couplings in the graphene layer at the $K$ and $K'$ points of the Brillouin zone. Thus, in principle, we need only consider the states at these high symmetry points in the bands of the magnetic layer. The neglect of dispersion in  the magnetic bands does not change the symmetry of the required states, and the hopping between graphene and the magnetic layer does not depend on the hopping within the magnetic layer. Hence, this approximation is sufficient to capture the general properties of the effective couplings in which we are interested. The geometry of the model is sketched in Fig.~\ref{fig: lattice}a.

The spin-orbit coupling splits the electronic levels of the magnetic layer into multiplets defined by the total angular momentum $\mathbf{J} = \mathbf{L}+\mathbf{S}$. We expand the couplings induced in the graphene layer in powers of momenta around the $K$ and $K'$ points. The leading terms are momentum-independent. The effective Hamiltonian at the $K$ and $K'$ points is characterized by three variables which can take two values each: sublattice, valley, and spin \cite{CGNG09}. Hence, the effective Hamiltonian can be written as an $8 \times 8$ matrix. This matrix can be expressed in terms of $2 \times 2$ Pauli matrices in the sublattice, valley, and spin subspaces, $\sigma_\mu , \tau_\mu$, and $s_\mu$, with  $\mu \in [0,1,2,3]$. The unperturbed Hamiltonian is then $
\mathcal{H}_0 = \epsilon_D \tau_0 \otimes \sigma_0 \otimes s_0$,
where $\epsilon_D$ is the Dirac energy, which we choose as the zero of energy. Here. we define matrices with $\mu = 0$ as identity matrices, which henceforth will be left implicit.

{\it Results.}
We now consider the effect of a ferromagnetic layer. Since the atoms on this layer are magnetized, we only consider the lowest-energy orbital for a given total angular momentum  for each magnetic atom; since it is magnetized, its angular momentum projection is fixed as well. This approximation holds when other orbitals lie at energies much further away from the Dirac energy of graphene, or, alternatively, when the hopping to graphene  from these magnetic orbitals is suppressed. In this limit, we  only allow the graphene electrons to hop to the  lowest-energy magnetic orbitals on the nearest-neighbor magnetic atoms. This modifies the low-energy band structure of graphene. In principle, these proximity-induced interaction terms can, to lowest order, be written as linear combinations of $\tau_\mu\otimes \sigma_\nu \otimes s_\upsilon$ that respect the underlying symmetries of the lattice. When the magnetization is out-of-plane, i.e., in the $z$-direction, the relevant symmetries of the Bloch wavefunctions at the $K$ and $K'$ points are 2D inversion, $\mathcal{I}$, and
$120^\circ$  rotation, $\mathcal{R}_\frac{2\pi}{3}$:
\begin{equation}
\begin{split}
\mathcal{I} &= \tau_x \otimes \sigma_x \otimes i s_z, \\
\mathcal{R}_\frac{2\pi}{3} &=  \left( \frac{1}{2}+i\frac{\sqrt{3}}{2} \tau_z \otimes \sigma_z \right) \otimes \left( \frac{1}{2}+i\frac{\sqrt{3}}{2} s_z  \right).\\
\end{split}
\end{equation}
All the terms which respect these two symmetries are listed in Table~\ref{tab: possible terms} (see the appendix for additional details).  Throughout this work, we neglect terms which contain $K$-$K'$ mixing elements, since intervalley scattering is negligible unless the bilayer system forms a superlattice of at least three graphene unit cells. We further classify the remaining terms according to their transformation under time reversal, defined by
\begin{equation}
\mathcal{T} = \tau_x \otimes is_y \mathcal{K},
\end{equation}
where $\mathcal{K}$ is the complex conjugation operator. 

When the magnetization is aligned in the plane of the substrate, we break rotational and inversion symmetries. However, when the magnetization points to a graphene carbon atom (e.g. along the $x$-direction) or points to the midpoint of the line joining two nearest-neighbor carbon atoms (e.g. along the $y$-direction), we still have mirror reflection symmetry in the plane perpendicular to the magnetization direction, as shown in Fig.~\ref{fig: lattice}c~and~d. In these two special orientations, the number of allowed terms is restricted by reflection symmetry. For instance, if the magnetization is aligned along the $x$-direction, then the lattice is invariant under reflection over the $y$-$z$ plane, and the corresponding Hamiltonian commutes with the mirror operator
\begin{equation}
\mathcal{M}_y = \sigma_x \otimes is_x.
\end{equation}
Similarly, if the magnetization is in the $y$-direction, then the lattice is invariant under reflection over the $x$-$z$ plane, and the corresponding Hamiltonian commutes with the mirror operator
\begin{equation}
\mathcal{M}_x = \tau_x \otimes is_y.
\end{equation}
All the terms which are consistent with these reflection symmetries are listed in Table~\ref{tab: possible terms}.

\begin{table}
\begin{center}
 \begin{tabular}{| c || c | c|} 
 \hline
 Magnetization & $\mathcal{T}$-symmetric & $\mathcal{T}$-antisymmetric  \\ 
 \hline \hline
 Out-of-plane, $z$ & \begin{tabular}{c}  $\tau_z \otimes \sigma_z \otimes s_z $ \\ $\sigma_y \otimes s_x-\tau_z \otimes \sigma_x \otimes s_y $  \end{tabular} & \begin{tabular}{c}  $ s_z $ \\ $\tau_z \otimes \sigma_z$   \end{tabular} \\ 
 \hline
 In-plane, $x$ &  \begin{tabular}{c}  $\sigma_x$ \\ $ \sigma_y \otimes s_z $ \\ $\tau_z \otimes s_x $ \\ $\tau_z \otimes \sigma_x \otimes s_x $  \\ $\tau_z \otimes \sigma_z \otimes s_y$   \end{tabular} &\begin{tabular}{c}  $s_x$  \\  $ \sigma_x \otimes s_x $   \\ $\sigma_z \otimes  s_z $ \\ $ \tau_z $ \\ $\tau_z \otimes \sigma_x $ \\ $ \tau_z \otimes \sigma_y \otimes s_z $  \end{tabular} \\ 
 \hline
 In-plane, $y$ & \begin{tabular}{c}  $\sigma_x$ \\ $ \sigma_z$ \\ $ \sigma_y \otimes s_y$  \\ $ \tau_z \otimes s_z$   \\ $ \tau_z \otimes \sigma_x \otimes s_z$ \\ $ \tau_z \otimes \sigma_z \otimes s_x$  \end{tabular} & \begin{tabular}{c}  $s_y$ \\ $ \sigma_y$ \\ $ \sigma_x \otimes s_y$ \\ $ \sigma_z \otimes s_y$  \\ $ \tau_z \otimes \sigma_y \otimes s_z$  \end{tabular}   \\ 
 \hline
\end{tabular}
\end{center}
\caption{Classification of possible proximity-induced interaction terms based on symmetries. The left column lists the direction of the magnetization, and each row lists all the terms which are consistent with the symmetries imposed by that magnetization direction. Terms in the middle column are even under time-inversion, and terms in the right column are odd. We have omitted terms which are related to each other by a 90$^\text{o}$-rotation in the spin Hilbert space about the magnetization axis.}
\label{tab: possible terms}
\end{table}

Based only on symmetry considerations, we have arrived at a set of possible proximity-induced interaction terms for three independent magnetization directions. We now construct simple microscopic tight-binding models that give us magnitude estimates of these terms. We consider a magnetic $d$-orbital with magnetization in a particular direction. The hopping between the graphene orbitals and the magnetic orbitals can be approximated by the Slater-Koster method \cite{SK54} in terms of two-center integrals $V_{pd\sigma}$ and $V_{pd\pi}$ that are exponentially suppressed by increasing distance between the centers of the orbitals. In the limit of weak coupling between the magnetic ions and the graphene carbon atoms, we can consider the effect of hopping between the two layers as a small perturbation on the wavefunction of graphene at the Dirac points. The first-order correction to the graphene wavefunction is found by summing the hopping orbitals over a graphene hexagon
\begin{equation}
\ket{\delta \psi_G} = \sum_{\tau, \sigma, s,j} \frac{t^j_s}{\epsilon_M} \ket{\psi^j_{\tau, \sigma, s}},
\end{equation}
where $\tau$, $\sigma$, $s$, denote the valley, sublattice, and  spin indices at the lattice site $j$ around the hexagon. The coefficient $t^j_s$ is the strength of the hopping from the carbon atom at site $j$ with spin $s$ to the magnetic orbital, $\epsilon_M$ is the energy of the magnetic orbital, and $\ket{\psi^j_{\tau, \sigma, s}}$ is the graphene orbital with the appropriate phase as defined in Fig.~\ref{fig: lattice}b. The effective Hamiltonian is found by projecting the full Hamiltonian onto the basis of the perturbed wavefunctions.

In the case of magnetization in the out-of-plane direction, we find the following  effective Hamiltonian 
\begin{equation}
\label{eq: effective Hamiltonian in the z-direction}
\begin{split}
\mathcal{H}_\text{eff} &= \lambda_\text{Z}s_z+ \lambda_\text{QH} \tau_z \otimes \sigma_z +\lambda_\text{QSH} \tau_z \otimes \sigma_z \otimes s_z\\
&+ \lambda_\text{SO} \left( \tau_z \otimes \sigma_x \otimes s_y-\sigma_y \otimes s_x\right).
\end{split}
\end{equation}
Here, the $s_z$ term is a Zeeman exchange, the $\tau_z \otimes \sigma_z$ term is the Haldane quantum orbital  Hall effect \cite{H88}, the $\tau_z \otimes \sigma_z \otimes s_z$ term is the Kane-Mele quantum spin Hall effect \cite{KM05}, and the $\left( \tau_z \otimes \sigma_x \otimes s_y-\sigma_y \otimes s_x\right)$ term is the Rashba spin-orbit coupling \cite{M06}. From this simple tight-binding model, we have found all four possible physically-distinct couplings that are allowed by symmetry. Separately, these terms have been studied and classified topologically. The Haldane term leads to the quantum anomalous Hall effect in the absence of external magnetic fields. The Kane-Mele term, in a system with time-reversal symmetry, gives rise to symmetry-protected topological edge states. The exchange term and the Rashba term independently do not give rise to gaps near the Dirac points. However, a combination of these two terms also leads to the quantum anomalous Hall effect \cite{TQ11,QW14}. Our perturbative microscopic model suggests that all of these terms will be present with the same order of magnitude as all of them are due to the same hopping processes from graphene to magnetic orbitals and back. Therefore, they must be considered simultaneously. For instance, for a fully spin-polarized magnetic orbital with $j = 5/2$ and $j_z=5/2$, the Rashba coupling term is extremely small, and all the other terms are approximately equal in magnitude. This is because the Rashba term describes spin-flip processes that are prohibited by having a spin-polarized orbital. To obtain the Rashba coupling, we should instead consider a magnetic orbital with non-maximal spin projection, such as a state with $j=5/2$ and $j_z = 3/2$. In this case, the tight-binding approach predicts that the Rashba term is of the same order of magnitude as the other three terms. 

\begin{figure}
\includegraphics[scale=0.35]{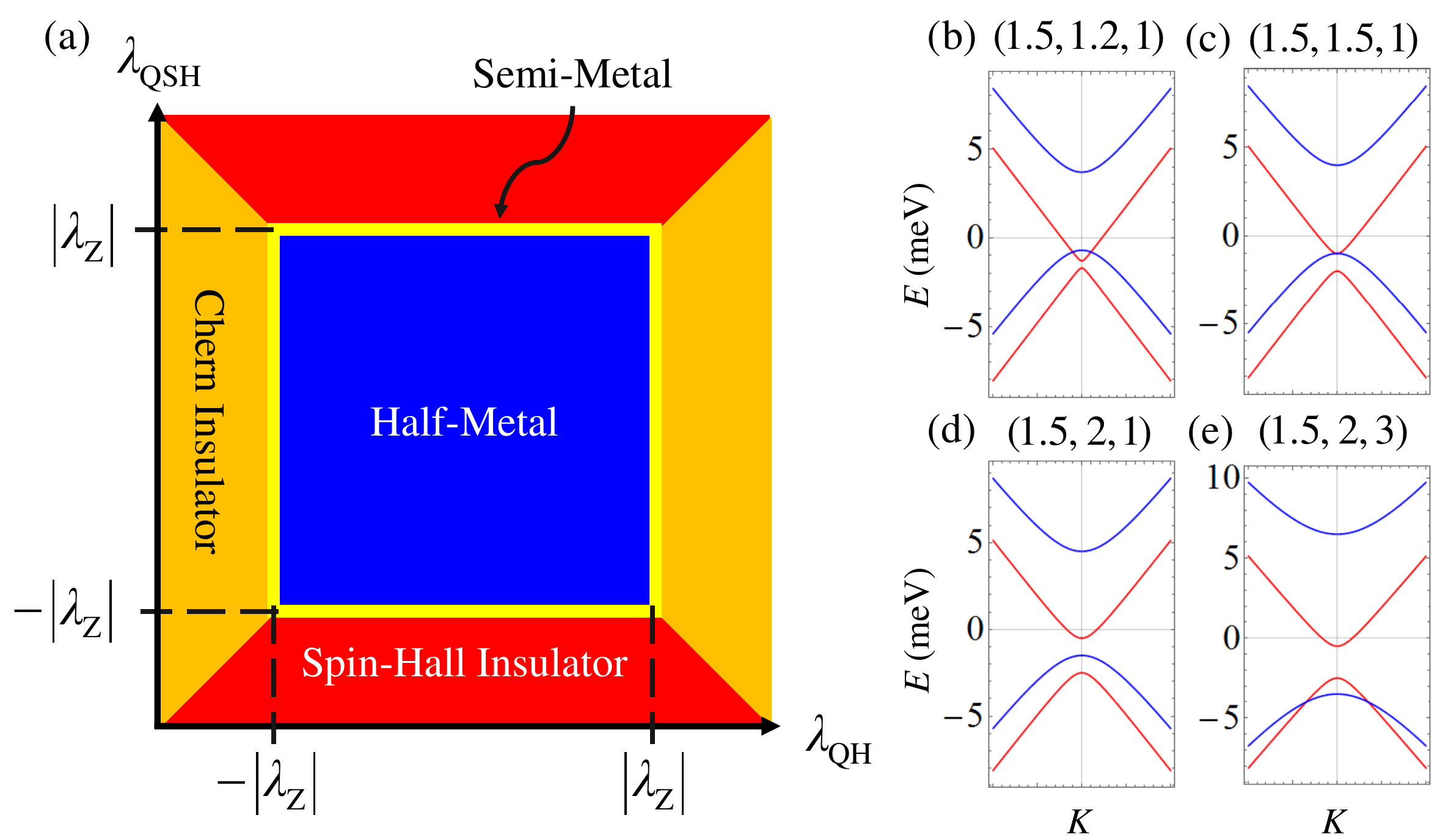}
\caption{(a) Phase diagram from proximity-induced interactions due to a spin-polarized magnetic orbital in the out-of-plane direction. (b)-(e) Representative band structures around the $K$ point for the half-metallic, semi-metallic, quantum Hall, and quantum spin Hall phases respectively. The coupling constants are given in triplets of the form $(\lambda_\text{Z},\lambda_\text{QH},\lambda_\text{QSH})$ in units of meV. Blue bands belong to up-spins, and red bands belong to down-spins.}
\label{fig: phasediagram1}
\end{figure}

These terms in combination give rise to a rich tapestry of topological features \cite{YXS11, M13,CVSCR14, SHHJ14, WW15}. For a spin-polarized orbital in the $z$-direction, we can induce a Zeeman splitting, a Haldane orbital-hopping, and a Kane-Mele spin-orbit coupling. When the Zeeman term dominates, $|\lambda_\text{Z}|>|\lambda_\text{QH}|$ and $|\lambda_\text{Z}|>|\lambda_\text{QSH}|$, we have a half-metal, meaning that if the Fermi energy is in the bulk band gap for one spin species, it lies inside the conduction band or the valence band of the other spin species. Consequently, electrons at the Fermi surface are spin-polarized, and can be harnessed for spintronic applications. Different spin species can be selected by adjusting the Fermi level or by changing the sign of $\lambda_\text{Z}$. Increasing the strength of the Haldane and the Kane-Mele terms, we enter the insulating phase where there is a common bulk gap for both spin species. Assuming the Fermi energy is in the gap, this phase can be classified according to the Chern numbers $\mathcal{C}_{\tau,s}$ for the valence bands of spins $s$ at the $\tau$ valleys
\begin{equation}
\mathcal{C}_{\tau,s} = \frac{1}{2\pi}\int d^2\mathbf{k}(\boldsymbol{\Omega}_{\tau,s})_z,
\end{equation} 
where the Berry curvature is given by
\begin{equation}
\boldsymbol{\Omega}_{\tau,s} = i \bra{\nabla_\mathbf{k} \psi_{\tau,s}}\times \ket{\nabla_\mathbf{k} \psi_{\tau,s}},
\end{equation}
and $\ket{ \psi_{\tau,s}}$ are valence-band eigenstates of the low-energy Hamiltonian around the $\tau$ valley. When an insulator has a non-zero Chern number, $\mathcal{C} =  \sum_\tau \left(\mathcal{C}_{\tau,\uparrow} + \mathcal{C}_{\tau,\downarrow}\right)$, we have a Chern insulator that hosts the quantum Hall effect with Hall conductance $C = e^2\mathcal{C}/h$. This occurs when the Haldane term dominates the Kane-Mele term, $| \lambda_\text{QSH}| < |\lambda_\text{QH} |$. Where the Chern number vanishes, we can define a different topological invariant called the spin Chern number $2\mathcal{C}_s =  \sum_\tau \left(\mathcal{C}_{\tau,\uparrow} - \mathcal{C}_{\tau,\downarrow}\right)$. When $\mathcal{C}_s =\pm 1$, this phase supports the quantum spin Hall effect where there are spin-polarized edge states. However, unlike in a time-reversal-invariant topological insulator, these edge states are not symmetry-protected, and can backscatter \cite{YXS11}. This phase occurs when the Kane-Mele term dominates over the Haldane term, $| \lambda_\text{QH}| < |\lambda_\text{QSH} |$. The cross-over boundary between the insulating phases and the half-metallic phases hosts a semi-metallic phase.  Our tight-binding model suggests that the strengths of all three terms are similar when we have a spin-polarized orbital for the magnetic atoms. This means all four phases of the diagram in Fig.~\ref{fig: phasediagram1} are  accessible via slight adjustments of the three coupling constants.
As discussed below, the value of the gap can be in the order of a few meV, so that extremely clean samples may be required to observe quantized edge conductances. On the other hand, each type of electronic structure gives rise to different distributions of Berry curvature in the Brillouin zone, and will modify in a different way the anomalous Hall effect \cite{Netal10} in the metallic regime.

When the magnetization is in-plane, many more coupling terms are allowed, since this configuration breaks both 2D inversion and rotation symmetries. However, at the two special alignments noted above, the Hamiltonian retains a mirror symmetry that restricts the number of possible terms. 
See the appendix for a detailed analysis of all terms.
As an example, when the magnetization is in the $x$-direction, we have eleven possible distinct couplings. Seven of these coupling terms commute with $s_x$, and thus conserve spin. These terms do not open gaps in the energy spectrum, but instead describe exchange interactions and pseudo-gauge potentials. As illustration, terms containing $\sigma_x$, e.g.,  $\sigma_x$, $\sigma_x \otimes s_x$ and $\tau_z \otimes \sigma_x$,  shift the Dirac points along the $k_x$-direction. On the other hand, terms containing $s_x$ such as $s_x$, $\tau_z \otimes s_x$ and $\tau_z \otimes \sigma_x \otimes s_x$ lift the spin degeneracy in the $x$-direction, acting as exchange interactions. When the magnetic orbital is not of maximal projection, spin-flip processes give rise to spin-projection non-conserving terms such as $\sigma_z \otimes s_y$, $\sigma_y \otimes s_z$, $\tau_z \otimes \sigma_z \otimes s_y$, and $\tau_z \otimes \sigma_y \otimes s_z$. Some of these terms open gaps in the energy spectrum. A similar structure can be found when the magnetization is aligned along the $y$-direction. As in the case of out-of-plane magnetization, we expect all of these coupling terms to be of similar order of magnitude. However, in addition to gap-opening and exchange terms, in the in-plane case, we also find pseudo-gauge potentials which can further modify spin-transport properties in graphene \cite{HHGB09, VKG10}. 

{\it Discussion and conclusions.}
We have presented an analysis of the general properties of the spin-dependent couplings induced in graphene by  proximity to a ferromagnetic insulator. A more quantitative analysis requires the knowledge of the band structure of the insulator, the alignment of the graphene layer with respect to the insulator, and, more crucially, the hopping between graphene and substrate orbitals. The last parameter is still poorly understood. The effective couplings within the graphene layer depend quadratically on this parameter. Theoretical estimates, mostly based on DFT calculations, for related systems give values in the $10^{-1} - 10^1$ meV range \cite{Wetal15,GF15,ZGFF16,Hetal17,KIF17,GF17,Y17}. Experimental values for these couplings also show a significant variety, although they tend to be smaller than the the theoretical predictions,  $\lesssim 0.1$ meV \cite{Aetal14,Wetal15,WLLPCCKZHH2016,Wetal16,Zetal17,Aetal17,Aetal17b,Setal17,KOVSP17,LKWW17}. The reason for this discrepancy is likely to be the uncertainty on the graphene-substrate distance, and the degree of commensurability between graphene and the substrate. The existance of levels in the magnetic substrate close to the Dirac energy of graphene may enhance significantly these couplings. We have analyzed here a few high symmetry orientations of the magnetization of the substrate, which we expect to be representative of the rich variety of regimes possible.

Irrespective of the imprecision in the value of the graphene-substrate couplings, our results lead to four main conclusions:

 i) The coupling of graphene to a magnetic substrate gives rise to a number of effective interactions beyond a simple exchange term. Some of them break time-reversal symmetry, and depend on the substrate's magnetization, and some others do not break time-reversal symmetry, and describe spin-orbit coupling terms.
 
 ii) A large fraction of the interactions mentioned above are of the same order of magnitude, as all of them are proportional to the square of the hopping between graphene and the substrate, divided by the energy difference between the states involved. 
 
 iii) The precise value of the couplings depends, among other factors, on the orientation of the magnetization of the substrate, leading to the possibility of modifying these interactions by changing the direction of the magnetization.
 
 iv) The interactions can open gaps at the Dirac energy of graphene. A rich phase diagram, which includes topological insulator and anomalous Hall insulator phases, emerges.

The above features lead to a rich number of regimes in the anomalous Hall effect typical of a ferromagnetic system. The three main contributions, intrinsic, skew scattering, and side jump \cite{Netal10}, are expected to depend on the various electronic structures described here, which also depend on the orientation of the magnetization. These topics will be discussed elsewhere.


\begin{acknowledgments}
We are thankful for helpful discussions with H. Ochoa and J. Fabian.
This work was supported by
funding from the European Union through the ERC Advanced
Grant NOVGRAPHENE through grant agreement
Nr. 290846, and from the European Commission under
the Graphene Flagship, contract CNECTICT-604391. VTP acknowledges financial support from the Marshall Aid Commemoration Commission. 
\end{acknowledgments}


\bibliography{paper_draft}

\appendix

\onecolumngrid
\newpage
\setcounter{equation}{0}
\setcounter{figure}{0}
\setcounter{table}{0}

\renewcommand{\theequation}{A\arabic{equation}}
\renewcommand{\thefigure}{A\arabic{figure}}
\renewcommand{\thetable}{A\Roman{table}}

\renewcommand{\bibnumfmt}[1]{[#1]}
\renewcommand{\citenumfont}[1]{#1}

\section{Tight-Binding Model}

In this section, we give details of the tight-binding model used in the main text. As mentioned there, we consider a bilayer system with a graphene bottom layer and a triangular-lattice ferromagnetic top layer, as shown in Fig.~\ref{fig: lattice} in the main text. Each carbon atom in the graphene layer has a free $p_z$ orbital onto which electrons can hop. We use a simple nearest-neighbor hopping model inside the graphene layer, where each carbon atom has three nearest neighbors at a distance $a$, to which its electron can hop. We shall concentrate on the low-energy properties, which
are dominated by the inequivalent $K$ and $K'$ points in the Brillouin zone. Here the electronic wavefunction is a combination of the localized orbital wavefunctions on the $A$ and $B$ sublattices with different spins. We can write the graphene wavefunction $\ket{\psi_G}$ in this basis as 
\begin{equation}
\ket{\psi_G} = \left( \begin{array}{ccccccccccccccc}
\psi_{K,A,s} & \psi_{K,A,s'} & \psi_{K,B,s} & \psi_{K,B,s'} &\psi_{K',A,s} & \psi_{K',A,s'} & \psi_{K',B,s} & \psi_{K',B,s'}  \end{array} \right).
\end{equation}
Here $K$ and $K'$ label the valley degree of freedom, $A$ and $B$ label the two sublattices, and $s$ and $s'$ are the  spin projections. The unperturbed Hamiltonian at these Dirac points is 
\begin{equation}
\mathcal{H}_0 = \epsilon_D \tau_0 \otimes \sigma_0 \otimes s_0,
\end{equation}
where $\epsilon_D$ is the Dirac energy, which we henceforth choose as zero. The $\tau_\mu$, $\sigma_\mu$, and $s_\mu$ Pauli matrices act on the valley, sublattice, and spin spaces respectively, and $\mu \in [0,1,2,3]$, where Pauli matrices with a zero index are the $2 \times 2$ identity matrix.

In considering  the ferromagnetic layer, we only include one energetically-favorable magnetized orbital per magnetic atom. We model this layer as a flat-band structure, ignoring the interactions inside the substrate that 
would allow electrons to hop in the magnetic material. Thus, electrons can only move by hopping to graphene atoms. We assume that the graphene electrons hop to nearest-neighbor magnetic orbitals only. In the limit of weak hopping, we can use lowest-order non-vanishing perturbation theory to find the effective Hamiltonian describing the change to the Dirac spectrum. 

To illustrate the procedure, we  consider the case where the magnetization is in the $z$-direction, perpendicular to the substrate. The simplest case is to assume a fully spin-polarized atomic orbital in the magnet. We consider a $d$ orbital with orbital angular momentum $l = 2$ and projection $l_z = 2$, and total angular momentum $j = 5/2$ and projection $j_z = 5/2$. The wavefunction of the magnetic orbital can then be written as 
\begin{equation}
\label{eq: j = 5/2, jz = 5/2}
\ket{\psi_M^{j_z=5/2}}= \ket{\psi_{x^2-y^2,\uparrow}}+i\ket{\psi_{xy,\uparrow}},
\end{equation}
where $\psi_{x^2-y^2,\uparrow}$ and $\psi_{xy,\uparrow}$ are real atomic $d$ orbitals with spatial symmetry indicated by the coordinate subscripts and spin indicated by $\uparrow$. In the absence of spin-orbit coupling, the orbital defined in Eq.~(\ref{eq: j = 5/2, jz = 5/2}) is degenerate with an orbital defined similarly but with the spins flipped. In what follows, we assume that there is strong spin-orbit coupling inside the magnetic ion that lifts this degeneracy. 

With the lattice cell defined in Fig.~\ref{fig: lattice}, each unit cell consists of three atoms, two graphene carbon atoms and one magnetic ion. This corresponds to three orbitals per unit cell, two $p_z$ orbitals from the carbon atoms, and one $d$ orbital from the magnetic ion. Without loss of generality, we set the energy of the $p_z$ orbitals to zero at the Dirac points, and denote the energy of the $d$ orbital as $\epsilon_{M}$. We only consider nearest-neighbor couplings. This means that an electron in a $d$ orbital can hop to six possible $p_z$ graphene orbitals, while an electron from a graphene orbital can hop to any of the three neighboring graphene orbitals or to any of the three neighboring magnetic orbitals.  Therefore, there are nine, $(3 \times 6)/2 = 9$, possible hopping coefficients per unit cell. 

The hopping between graphene orbitals and magnetic orbitals can be approximated by the Slater-Koster (SK) method \cite{SK54} in terms of two-center integrals that depend on the distance between centers of the orbitals.  The coupling terms also depend on the relative orientation of the orbitals. Let us place the magnetic atom at $\mathbf{r}_M = (0,0, d_M)$, where $d_M$ is the vertical distance between the graphene plane and the substrate, and the six surrounding graphene atoms are at $\mathbf{r}^j_G = a(\cos[j\pi/3],\sin[j\pi/3],0)$, where $j = 0,1,...,5$, and $a$ is the distance between adjacent graphene atoms. The relevant SK parameters in this case are
\begin{equation}
\label{eq: SK paramters 1}
\begin{aligned}
E_{z,x^2-y^2}^j &= \frac{\sqrt{3}}{2}n_j(l_j^2-m_j^2)V_{pd\sigma}-n_j(l_j^2-m_j^2)V_{pd\pi}, \\
E_{z,xy}^j &= \sqrt{3}l_jm_jn_jV_{pd\sigma}-2l_jm_jn_j V_{pd\pi},
\end{aligned}
\end{equation}
where $V_{pd\sigma}$ and $V_{pd\pi}$ are the two-center integrals, and $l_j$, $m_j$ and $n_j$ are direction cosines of $\mathbf{r}_M-\mathbf{r}_G^j$ defined by 
\begin{equation}
\begin{aligned}
l_j &= \sqrt{1-n^2}\cos\left(\frac{\pi j}{3}\right),\\
m_j &= \sqrt{1-n^2}\sin\left(\frac{\pi j}{3}\right),\\
n_j &= n = \frac{d_M}{\sqrt{a^2+d_M^2}}.
\end{aligned}
\end{equation}
Since the magnetic orbital is spin-polarized, only spin-up graphene electrons are allowed to hop into this $d$ orbital. Putting in the spin index, the hoppings terms are just
\begin{equation}
\label{eq: hopping term for d orbital}
t_{M,s}^{j} =
  \begin{cases}
    E_{z,x^2-y^2}^j+i E_{z,xy}^j = \left(n-n^3 \right) \left(\frac{\sqrt{3}}{2}V_{pd\sigma}-V_{pd\pi}  \right)e^{\frac{2\pi i}{3}j}= t_Me^{\frac{2\pi i}{3}j} & s = \uparrow \\
    0 & s = \downarrow 
  \end{cases}.
\end{equation}
In essence, we can now augment the graphene tight-binding model to include the $d$ orbitals of the magnetic atoms with on-site energy $\epsilon_M$ and nearest-neighbor hoppings defined by Eq.~(\ref{eq: hopping term for d orbital}).

\begin{figure}
\includegraphics[scale=0.3]{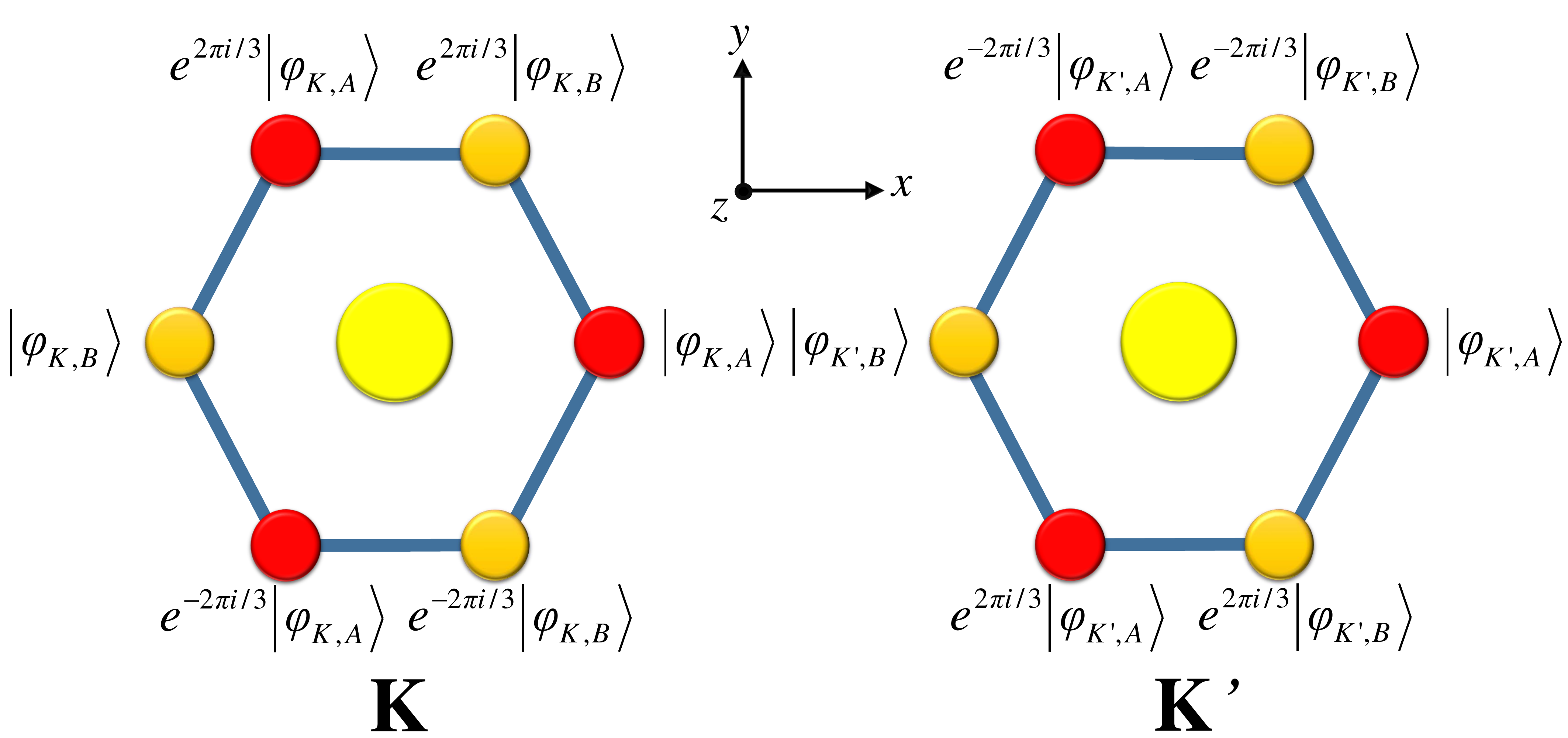}
\caption{Electron wavefunction of the graphene lattice at the $K$ and $K'$ Dirac points with different phases around the hexagon. $\ket{\psi_{K,A}}$, $\ket{\psi_{K,B}}$, $\ket{\psi_{K',A}}$, and $\ket{\psi_{K',B}}$ are the orbital wavefunctions of the $A$ and $B$ sublattices. }
\label{fig: graphene wavefunction}
\end{figure}

In the limit of weak hopping between the magnetic atoms and the graphene atoms, $t_M^2 \ll t_G \epsilon_M$, where $t_G$ is the hopping magnitude between carbon orbitals, we can consider the effect of hopping between the two layers as a small perturbation on the wavefunction of graphene at the Dirac points. To first order, the correction to the graphene wavefunction is 
\begin{equation}
\ket{\delta \psi_G} = \sum_{ \tau \in \lbrace K, K' \rbrace, s \in \lbrace \uparrow, \downarrow\rbrace, j} \frac{t_{M,s}^{j}}{\epsilon_{M,s}}|\phi_{ \tau,s}^{j} \rangle,
\end{equation}
where $\phi_{\sigma, \tau}^j$ is the atomic orbital of graphene at site $j$ in the $\tau \in \lbrace K, K'\rbrace$ point of the Brillouin zone with spin $s \in \lbrace \uparrow, \downarrow \rbrace$ with the appropriate phase shown in Fig.~\ref{fig: graphene wavefunction}. Summing over the atomic orbitals with the correct phases, we find that the phases interfere destructively among the $B$ graphene orbitals and constructively among the $A$ graphene orbitals in the $K$ point. Conversely, the graphene orbitals interfere destructively among the $A$ graphene orbitals and constructively among the $B$ graphene orbitals in the $K'$ point. That is, the perturbed wavefunction is 
\begin{equation}
\ket{\delta \psi_G} = \frac{3t_M}{\epsilon_{M}} \left(|\phi_{ K, A, \uparrow} \rangle + |\phi_{ K', B, \uparrow} \rangle \right).
\end{equation}
Neglecting intervalley terms, the effective Hamiltonian is 
\begin{equation}
\label{eq: Hamiltonian for jz = 5/2}
\mathcal{H}_\text{eff} = \frac{9t_M^2}{\epsilon_M}\left( c^\dagger_{K,A, \uparrow} c_{K,A, \uparrow} +c^\dagger_{K',B, \uparrow} c_{K',B, \uparrow}  \right),
\end{equation}
where $c^\dagger_{\tau, \sigma, s}$ and $c_{\tau, \sigma, s}$ are creation and annihilation operators that create or destroy an orbital at the $\tau$ valley, $\sigma$ sublattice, and with spin $s$. Writing Eq.~(\ref{eq: Hamiltonian for jz = 5/2}) in the basis of Pauli matrices, we have
\begin{equation}
\label{eq: Hamiltonian with Pauli matrices for jz = 5/2}
\mathcal{H}_\text{eff} = \frac{9t_M^2}{4\epsilon_M}\left( \mathbb{1}+\tau_z\otimes\sigma_z \right)\left( \mathbb{1}+s_z \right) = \frac{9t_M^2}{4\epsilon_M}\left( \mathbb{1}+ s_z + \tau_z\otimes\sigma_z  +\tau_z\otimes\sigma_z  \otimes s_z \right).
\end{equation}
In considering just a $d$ orbital with maximal perpendicular angular momentum, we find that several coupling terms are possible. These terms modify the low-energy spectrum near the Dirac points of graphene. 


We can consider the effect of coupling to orbitals which are not fully spin-polarized. The relative strengths of the four operators on the right-hand side of Eq.~(\ref{eq: Hamiltonian with Pauli matrices for jz = 5/2}) will no longer be equal. Furthermore, new terms may also be possible. Orbitals which do not have maximal angular momentum projection combine wavefunctions with up and down spins, and can give these new effects. For instance, we consider the same $d$ orbital as before,  with $j = 5/2$ but now take $j_z = 3/2$ with wavefunction
\begin{equation}
\label{eq: wavefunction for jz=3/2}
\ket{\psi_M^{j_z=3/2}} = \frac{1}{\sqrt{5}} \left( \ket{\psi_{x^2-y^2,\downarrow}}+i\ket{\psi_{xy,\downarrow}} \right)+\frac{2}{\sqrt{5}} \left( \ket{\psi_{xz,\uparrow}}+i\ket{\psi_{yz,\uparrow}} \right).
\end{equation}
As we can see from Eq.~(\ref{eq: wavefunction for jz=3/2}), the spin-orbit coupling combines wavefunctions with opposite spins and different orbital angular momenta. In addition to the SK parameters in Eq.~(\ref{eq: SK paramters 1}), there are two more relevant parameters
\begin{equation}
\label{eq: SK parameters 2}
\begin{aligned}
E_{z,xz}^j &=  \sqrt{3} l_jn_j^2V_{pd\sigma}+l_j(1-2n_j^2)V_{pd\pi},\\
E_{z,yz}^j &= \sqrt{3} m_jn_j^2V_{pd\sigma}+m_j(1-2n_j^2)V_{pd\pi}.
\end{aligned}
\end{equation}
The hopping coefficients are then
\begin{equation}
\label{eq: hopping term for d orbital2}
t_{M,s}^{j} =
  \begin{cases}
    t_{M,\uparrow}e^{\frac{\pi i}{3}j} & s = \uparrow \\
     t_{M,\downarrow}e^{\frac{2\pi i}{3}j} & s = \downarrow 
  \end{cases},
\end{equation}
where the magnitudes are given by 
\begin{equation}
\begin{aligned}
 t_{M,\downarrow} & =\frac{n-n^3}{\sqrt{5}}  \left(\frac{\sqrt{3}}{2}V_{pd\sigma}-V_{pd\pi}  \right),\\
 t_{M,\uparrow}&= \frac{2}{\sqrt{5}}\sqrt{1-n^2} \left( n^2 \left(\sqrt{3}V_{pd\sigma} - 2V_{pd\pi} \right)+V_{pd\pi} \right).
\end{aligned} 
\end{equation}
The perturbed wavefunction is 
\begin{equation}
\ket{\delta \psi_G} = \frac{3 t_{M, \downarrow}}{\epsilon_M} \left( \ket{\phi_{ K, A, \downarrow}} + \ket{\phi_{ K', B, \downarrow}}\right)+ \frac{3 t_{M, \uparrow}}{\epsilon_M} \left( \ket{\phi_{ K', A, \uparrow}} - \ket{\phi_{ K, B, \uparrow}}\right).
\end{equation}
The effective Hamiltonian is 
\begin{equation}
\begin{split}
\mathcal{H}_\text{eff} &= \frac{9t_{M,\downarrow}^2}{\epsilon_M}\left( c^\dagger_{K,A,\downarrow}c_{K,A,\downarrow}+c^\dagger_{K',B,\downarrow}c_{K',B,\downarrow} \right)+\frac{9t_{M,\uparrow}^2}{\epsilon_M}\left( c^\dagger_{K',A,\uparrow}c_{K',A,\uparrow}+c^\dagger_{K,B,\uparrow}c_{K,B,\uparrow} \right) \\
&+ \frac{9 t_{M,\downarrow}t_{M,\uparrow}}{\epsilon_M} \left( c^\dagger_{K',B,\downarrow} c_{K',A, \uparrow} +c^\dagger_{K',A, \uparrow} c_{K',B, \downarrow} - c^\dagger_{K,B,\uparrow} c_{K, A, \downarrow} - c^\dagger_{K, A, \downarrow} c_{K, B, \uparrow}  \right).
\end{split}
\end{equation}
In terms of Pauli matrices, the effective Hamiltonian is 
\begin{equation}
\label{eq: Hamiltonian for jz=3/2 with Pauli matrices}
\mathcal{H}_\text{eff} = \lambda_1 \left( \mathbb{1}-\tau_z \otimes \sigma_z \otimes s_z \right) +\lambda_2 \left( s_z- \tau_z \otimes \sigma_z\right) - \lambda_3 \left( \sigma_y \otimes s_y+\tau_z\otimes \sigma_x \otimes s_x \right),
\end{equation}
where 
\begin{equation}
\begin{split}
\lambda_1 = \frac{9 \left(t_{M,\downarrow}^2+t_{M,\uparrow}^2 \right)}{4\epsilon_M}, \quad \lambda_2 = \frac{9 \left(t_{M,\uparrow}^2-t_{M,\downarrow}^2 \right)}{4\epsilon_M}, \quad \text{and} \quad \lambda_3  = \frac{9 t_{M,\downarrow}t_{M,\uparrow}}{2 \epsilon_M}.
\end{split}
\end{equation}
The last term in Eq.~(\ref{eq: Hamiltonian for jz=3/2 with Pauli matrices}), upon a $90^\text{o}$-rotation in spin space about the $z$-axis, is just the Rashba coupling, $\tau_z\otimes \sigma_x \otimes s_y-\sigma_y \otimes s_x$. In this more familiar notation, the Hamiltonian is
\begin{equation}
\label{eq: Hamiltonian for jz=3/2 with Pauli matrices 2}
\mathcal{H}_\text{eff} = \lambda_1 \left( \mathbb{1}-\tau_z \otimes \sigma_z \otimes s_z \right) +\lambda_2 \left( s_z- \tau_z \otimes \sigma_z\right) + \lambda_3 \left( \tau_z\otimes \sigma_x \otimes s_y-\sigma_y \otimes s_x \right).
\end{equation}
The terms in Eq.~(\ref{eq: Hamiltonian for jz=3/2 with Pauli matrices 2}) are all the terms that are consistent with the spatial symmetries of the lattice.  With this simple microscopic model, we can estimate the relative contributions of the different terms.

The procedure above can likewise be used to find the coupling terms when the magnetization is in-plane. In general, we can write the effective Hamiltonian when the magnetization is in the $x$-direction as 
\begin{equation}
\begin{split}
\mathcal{H}_\text{eff} &= \xi_1 \sigma_x+\xi_2 s_x + \xi_3 \sigma_x \otimes s_x + \xi_4 \tau_z + \xi_5 \tau_z \otimes \sigma_x+\xi_6 \tau_z \otimes s_x + \xi_7 \tau_z \otimes \sigma_x \otimes s_x \\
&+ \xi_8 \sigma_z \otimes s_y + \xi_9 \sigma_y \otimes s_z + \xi_{10} \tau_z \otimes \sigma_z \otimes s_y + \xi_{11} \tau_z \otimes  \sigma_y \otimes s_z,
\end{split}
\end{equation}
where $\xi_i$ are coupling constants, for $i = 1,...,11$. The first seven terms commute with $s_x$, and therefore conserve spin in the $x$-direction. When we have spin polarization, only these seven terms are allowed. The remaining four terms mix the spins, and therefore, can only exist if we consider orbitals with non-maximal spin projection. We analyze these terms one-by-one. For the seven spin-conserving terms, we have exchange effects and pseudo-gauge potentials, which may be valley-dependent and spin-dependent. These terms shift the position of the Dirac points at the $K$ and $K'$ points and break the degeneracy of the right and left spins. Of the four terms that mix the spins, two terms behave as pseudo-gauge fields, and two terms open gaps in the energy spectrum. This is summarized in Table~\ref{tab: possible terms x direction}.

We now study the case where the magnetization is in the $y$-direction. In general, the Hamiltonian is 
\begin{equation}
\begin{split}
\mathcal{H}_\text{eff} &= \chi_1 \sigma_x + \chi_2 \sigma_y + \chi_3 \sigma_z + \chi_4 s_y  + \chi_5  \sigma_x \otimes s_y + \chi_6 \sigma_y \otimes s_y + \chi_7 \sigma_z \otimes s_y \\
&+ \chi_8 \tau_z \otimes s_z + \chi_9 \tau_z \otimes \sigma_x \otimes s_z + \chi_{10} \tau_z \otimes \sigma_y \otimes s_z + \chi_{11} \tau_z \otimes \sigma_z \otimes s_x,
\end{split}
\end{equation}
where $\chi_i$ are coupling constants, for $i = 1,...,11$. The first seven terms exist when we have spin polarization. The remaining four terms involve spin-flip processes, and can only be non-zero if we include non-maximally-magnetized orbitals. Furthermore, in our tight-binding model, $\chi_3 = \chi_7 = 0$ because these terms correspond to staggered potentials that require sublattice-symmetry breaking. Let us now analyze these terms one-by-one as before. Only terms which contain $\sigma_z$ open gaps in the energy spectrum. The other terms correspond to either exchange terms or pseudo-gauge potentials, which can dependent on spin and valley. If we ignore the staggered potentials, then the only remaining term that can open a gap is $\tau_z \otimes \sigma_z \otimes s_x$. The complete list of terms is summarized in Table~\ref{tab: possible terms y direction}.

\begin{table}[htb]
 \caption{Effect of coupling terms on the spectrum of graphene when the magnetization is in the $x$-direction. We denote the unperturbed spectrum $\epsilon(k) = v_Fk$, and the valley index $\tau = \pm 1$. When the Hamiltonian conserves $s_x$, we label the spectrum by the right and left spins $\rightarrow$ and $\leftarrow$. Otherwise, we label the states as $(1)$ and $(2)$. Of the eleven terms, only two terms open gaps in the energy spectrum.}
\label{tab: possible terms x direction}
\begin{center}
\begin{tabular}{| p{2.5cm} | p{6cm} | p{9cm} |} 
 \hline
 Term & Spectrum & Description  \\ 
 \hline \hline
 $ \xi_1 \sigma_x $ & $E_{\pm, \leftrightarrows} = \pm v_F\sqrt{k_y^2+\left(\tau k_x + \xi_1/v_F \right)^2}$ & Pseudo-gauge potential that shifts the Dirac point along the $k_x$-axis in opposite directions for the two valleys. This does not open a gap. \\
 \hline
 $ \xi_2 s_x $ & \begin{tabular}{l}  $E_{\pm, \rightarrow} = \xi_2 \pm \epsilon(k) $ \\ $E_{\pm, \leftarrow} = -\xi_2 \pm \epsilon(k) $  \end{tabular}  & Exchange potential that breaks the degeneracy between the right and left spins. This does not open a gap. \\
 \hline
 $ \xi_3 \sigma_x \otimes s_x $ &  \begin{tabular}{l}  $E_{\pm, \rightarrow} = \pm v_F\sqrt{k_y^2+\left(\tau k_x - \xi_3/v_F \right)^2} $ \\ $E_{\pm, \leftarrow} = \pm v_F\sqrt{k_y^2+\left(\tau k_x + \xi_3/v_F \right)^2}$  \end{tabular} & Spin-dependent pseudo-gauge potential that shifts the Dirac point along the $k_x$-axis in opposite directions for the two valleys. This does not open  a gap. \\
 \hline
 $ \xi_4 \tau_z $ & $E_{\pm, \leftrightarrows} = \tau \xi_4 \pm \epsilon(k)$  & Valley exchange term that breaks the degeneracy between the $K$ and $K'$ points. This does not open a gap.\\
 \hline
 $ \xi_5 \tau_z \otimes \sigma_x $ & $E_{\pm, \leftrightarrows} = \pm v_F\sqrt{k_y^2+\left( k_x + \xi_5/v_F \right)^2}$ & Pseudo-gauge potential that shifts the Dirac point along the $k_x$-axis in same direction for the two valleys. This does not open a gap.  \\
 \hline
 $ \xi_6 \tau_z \otimes s_x  $ & \begin{tabular}{l}  $E_{\pm, \rightarrow} = \tau \xi_6 \pm \epsilon(k) $ \\ $E_{\pm, \leftarrow} = -\tau \xi_6 \pm \epsilon(k) $  \end{tabular}  & Valley-dependent exchange potential that breaks the degeneracy between the right and left spins. This does not open a gap. \\
 \hline
 $ \xi_7 \tau_z \otimes \sigma_x \otimes s_x $ & \begin{tabular}{l}  $E_{\pm, \rightarrow} = \pm v_F\sqrt{k_y^2+\left( k_x - \xi_7/v_F \right)^2} $ \\ $E_{\pm, \leftarrow} = \pm v_F\sqrt{k_y^2+\left( k_x + \xi_7/v_F \right)^2}$  \end{tabular} & Spin-dependent pseudo-gauge potential that shifts the Dirac point along the $k_x$-axis in the same direction for the two valleys. This does not open a gap. \\
 \hline
 $ \xi_8 \sigma_z \otimes s_y  $ & $E_{\pm,(1,2)} = \pm \sqrt{\xi_8^2+\epsilon(k)^2}$& Sublattice-symmetry-breaking term that opens a gap of magnitude $2|\xi_8|$. \\
 \hline
 $ \xi_9 \sigma_y \otimes s_z $ & \begin{tabular}{l}  $E_{\pm, (1)} = \pm v_F \sqrt{(k_y+\xi_9/v_F)^2+k_x^2} $ \\ $E_{\pm, (2)} = \pm v_F \sqrt{(k_y-\xi_9/v_F)^2+k_x^2} $  \end{tabular} & Pseudo-gauge potential that shifts the Dirac point along the $k_y$-axis in the same direction for the two valleys, and also mixes the spins. This does not open a gap.  \\
 \hline
 $ \xi_{10} \tau_z \otimes \sigma_z \otimes s_y  $ & $E_{\pm,(1,2)} = \pm \sqrt{\xi_{10}^2+\epsilon(k)^2}$ & Valley-dependent sublattice-symmetry-breaking term that opens a gap of magnitude $2|\xi_{10}|$. \\
 \hline
 $ \xi_{11} \tau_z \otimes  \sigma_y \otimes s_z  $ &  \begin{tabular}{l}  $E_{\pm, (1)} = \pm v_F\sqrt{k_x^2+\left(\tau k_y + \xi_{11}/v_F \right)^2} $ \\ $E_{\pm, (2)} = \pm v_F\sqrt{k_x^2+\left(\tau k_y - \xi_{11}/v_F \right)^2}$  \end{tabular} & Pseudo-gauge potential that shifts the Dirac point along the $k_y$-axis in opposite directions for the two valleys. This does not open a gap. \\
 \hline
\end{tabular}
\end{center}

\end{table}

\begin{table}
\caption{Effect of the induced  couplings on the spectrum of graphene when the magnetization is in the $y$-direction. We denote the unperturbed spectrum $\epsilon(k) = v_Fk$, and the valley index $\tau = \pm 1$. When the Hamiltonian conserves $s_y$, we label the spectrum by the spin-eigenstates denoted as $\nearrow$ and $\swarrow$. Otherwise, we label the spectrum by $(1)$ and $(2)$. Of the eleven terms, three terms open gaps in the energy spectrum.}
\label{tab: possible terms y direction}
\begin{center}
 \begin{tabular}{| p{2.5cm} | p{6cm} | p{9cm} |} 
 \hline
 Term & Spectrum & Description  \\ 
 \hline \hline
 $ \chi_1 \sigma_x $ & $E_{\pm, \nearrow\swarrow} = \pm v_F\sqrt{k_y^2+\left(\tau k_x + \chi_1/v_F \right)^2}$ & Pseudo-gauge potential that shifts the Dirac point along the $k_x$-axis in opposite directions for the two valleys. This does not open a gap. \\
 \hline
 $ \chi_2 \sigma_y $ & $E_{\pm, \nearrow\swarrow} = \pm v_F\sqrt{k_x^2+\left( k_y + \chi_2/v_F \right)^2}$  & Pseudo-gauge potential that shifts the Dirac point along the $k_y$-axis in the same direction for the two valleys. This does not open a gap. \\
 \hline
 $  \chi_3 \sigma_z $ &  \begin{tabular}{l}  $E_{\pm, \nearrow\swarrow} = \pm \sqrt{\epsilon(k)^2+\chi_3^2}$  \end{tabular} & Mass term that breaks sublattice symmetry, and opens a gap of magnitude $2|\chi_3|$. \\
 \hline
 $ \chi_4 s_y $ &  \begin{tabular}{l} $E_{\pm, \nearrow} = \chi_4 \pm \epsilon(k) $ \\ $E_{\pm, \swarrow} = -\chi_4 \pm \epsilon(k) $  \end{tabular}  & Exchange potential that breaks the degeneracy between the in and out spins. This does not open a gap.\\
 \hline
 $ \chi_5  \sigma_x \otimes s_y $ & \begin{tabular}{l} $E_{\pm, \nearrow} = \pm \sqrt{k_y^2+\left( \tau k_x - \chi_5/v_F\right)^2} $ \\ $E_{\pm, \swarrow} = \pm \sqrt{k_y^2+\left( \tau k_x + \chi_5/v_F\right)^2} $  \end{tabular} & Spin-dependent pseudo-gauge potential that shifts the Dirac point along the $k_x$-axis in opposite directions for the two valleys. This does not open a gap.  \\
 \hline
 $  \chi_6 \sigma_y \otimes s_y   $ & \begin{tabular}{l} $E_{\pm, \nearrow} = \pm \sqrt{k_x^2+\left(  k_y - \chi_6/v_F\right)^2} $ \\ $E_{\pm, \swarrow} = \pm \sqrt{k_c^2+\left(  k_y + \chi_6/v_F\right)^2} $  \end{tabular}  & Spin-dependent pseudo-gauge potential that shifts the Dirac point along the $k_y$-axis in the same direction for the two valleys. This does not open a gap. \\
 \hline
 $ \chi_7 \sigma_z \otimes s_y $ &  \begin{tabular}{l}  $E_{\pm, \nearrow\swarrow} = \pm \sqrt{\epsilon(k)^2+\chi_7^2}$  \end{tabular} & Spin-dependent term that breaks sublattice symmetry, and opens a gap of  magnitude $2|\chi_7|$. \\
 \hline
 $ \chi_8 \tau_z \otimes s_z  $ &\begin{tabular}{l}  $E_{\pm, (1)} = \tau \chi_8 \pm \epsilon(k) $ \\ $E_{\pm, (2)} = -\tau \chi_8 \pm \epsilon(k) $  \end{tabular}  & Valley-dependent exchange-like potential that breaks the degeneracy between the in and out spins. This does not open a gap.  \\
 \hline
 $ \chi_9 \tau_z \otimes \sigma_x \otimes s_z $ & \begin{tabular}{l}  $E_{\pm, (1)} = \pm v_F \sqrt{(k_x+\chi_9/v_F)^2+k_y^2} $ \\ $E_{\pm, (2)} = \pm v_F \sqrt{(k_x-\chi_9/v_F)^2+k_y^2} $  \end{tabular} & Pseudo-gauge potential that shifts the Dirac point along the $k_x$-axis in the same direction for the two valleys. This does not open a gap.  \\
 \hline
 $ \chi_{10} \tau_z \otimes \sigma_y \otimes s_z $ & \begin{tabular}{l}  $E_{\pm, (1)} = \pm v_F \sqrt{(\tau k_y+\chi_{10}/v_F)^2+k_x^2} $ \\ $E_{\pm, (2)} = \pm v_F \sqrt{(\tau k_y-\chi_{10}/v_F)^2+k_x^2} $  \end{tabular} & Pseudo-gauge potential that shifts the Dirac point along the $k_y$-axis in opposite directions for the two valleys. This does not open a gap.  \\
 \hline
 $  \chi_{11} \tau_z \otimes \sigma_z \otimes s_x $ &  \begin{tabular}{l}  $E_{\pm, \nearrow\swarrow} = \pm \sqrt{\epsilon(k)^2+\chi_{11}^2}$  \end{tabular} & This term opens a gap of magnitude $2|\chi_{11}|$. \\
 \hline
\end{tabular}
\end{center}

\end{table}

\end{document}